# An Intelligent Deterministic Scheduling Method for Ultra-Low Latency Communication in Edge Enabled Industrial Internet of Things

Yinzhi Lu, Liu Yang, Simon X. Yang, *Senior Member*, *IEEE*, Qiaozhi Hua, Arun Kumar Sangaiah, *Member*, *IEEE*, Tan Guo, *Member*, *IEEE*, and Keping Yu, *Member*, *IEEE*

*Abstract*—Edge enabled Industrial Internet of Things (IIoT) platform is of great significance to accelerate the development of smart industry. However, with the dramatic increase in real-time IIoT applications, it is a great challenge to support fast response time, low latency, and efficient bandwidth utilization. To address this issue, Time Sensitive Network (TSN) is recently researched to realize low latency communication via deterministic scheduling. To the best of our knowledge, the combinability of multiple flows, which can significantly affect the scheduling performance, has never been systematically analyzed before. In this article, we first analyze the combinability problem. Then a non-collision theory based deterministic scheduling (NDS) method is proposed to achieve ultra-low latency communication for the time-sensitive flows. Moreover, to improve bandwidth utilization, a dynamic queue scheduling (DQS) method is presented for the best-effort flows. Experiment results demonstrate that NDS/DQS can well support deterministic ultra-low latency services and guarantee efficient bandwidth utilization.

## I. INTRODUCTION

WITH the rapid development of Industrial Internet of Things (IIoT), artificial intelligence (AI), and edge computing, a new revolution of Industry 4.0 and beyond is now in progress, for the purpose of realizing smart industry [1]. With this trend, edge enabled IIoT platform is of

Manuscript received January 25, 2022; revised April 20 and May 31, 2022; accepted June 15, 2022. This work was supported in part by the National Natural Science Foundation of China under Grant 61801072, in part by the Science and Technology Research Program of Chongqing Municipal Education Commission under Grant KJQN202000641, in part by the Natural Science Foundation of Chongqing under Grant cstc2020jcyj-msxmX0636, in part by the Key Scientific and Technological Innovation Project for 'Chengdu-Chongqing Double City Economic Circle' under Grant KJCXZD2020025, in part by the Japan Society for the Promotion of Science (JSPS) Grants-in-Aid for Scientific Research (KAKENHI) under Grants JP18K18044 and JP21K17736, and in part by the Natural Science Foundation of Hubei Province under Grant 2021CFB156. (*Corresponding authors*: *Qiaozhi Hua; Arun Kumar Sangaiah*.)

Yinzhi Lu, Liu Yang, and Tan Guo are with the School of Communication and Information Engineering, Chongqing University of Posts and Telecommunications, Chongqing 400065, China (e-mail:henanluyinzhi@163.com;yangliu@cqupt.edu.cn; guot@cqupt.edu.cn;).

Simon X. Yang is with the Advanced Robotics and Intelligent Systems Laboratory, School of Engineering, University of Guelph, Guelph, ON N1G2W1, Canada (e-mail: syang@uoguelph.ca).

Qiaozhi Hua is with the Computer School, Hubei University of Arts and Science, Xiangyang 441000, China (e-mail: 11722@hbuas.edu.cn).

Arun Kumar Sangaiah is with the National Yunlin University of Science and Technology, Taiwan (e-mail: aksangaiah@ieee.org).

Keping Yu is with the Graduate School of Science and Engineering, Hosei University, Tokyo 184-8584, Japan, and with the RIKEN Center for Advanced Intelligence Project, RIKEN, Tokyo 103-0027, Japan (email: keping.yu@ieee.org).

great significance to accelerate the development of intelligent applications in industry, such as industrial surveillance, smart manufacturing, and predictive maintenance. However, due to the heterogeneity and limited computing and communication capabilities of edge nodes, the rapid increase of real-time IIoT applications makes edge intelligence face a great challenge in terms of fast response, low latency, and efficient bandwidth utilization [2]. Therefore, how to satisfy rigorous requirements of IIoT systems is a significant problem to be solved.

Time Sensitive Network (TSN) is a promising technology that guarantees high real-time latency requirements for edge enabled IIoT [1], [3]. As an enhancement to Ethernet, TSN has recently attracted many researchers to explore its applications, for the purpose of assuring bounded low latency and jitters for time-sensitive flows and providing services for best-effort traffics [4]. Moreover, that TSN can help accelerating the revolution of IIoT to next generation operational efficiency and computing connectivity has reached a consensus. IEEE TSN Task Group has defined a whole range of 802.1 sub-standards, which involve the clock synchronization, frame preemption, network management, and scheduled traffic enhancement [5]. In particular, IEEE 802.1Qbv Time Aware Shaper (TAS) has been known as a key technique to guarantee the deterministic low latency requirements via time-triggered communication that follows static schedules [4], [5].

In the past few years, some scheduling methods have been presented to get the global static schedules, with the aim of guaranteeing deterministic end-to-end latency for time-critical traffic in TSN [3], [6]. Queuing theory has been adopted to analyze the end-to-end transmission latency, and an analysis of the worst-case latency for time-sensitive flows has been made using network calculus [7], [8]. Whereas, one of the major drawbacks of the two methods is that the statistical analyses of the transmission latency are based on the hypothetical arrival patterns of flows. Therefore, they are not suitable for latency analyzing in deterministic networks. In literatures [9] and [10], the scheduling of time-sensitive flows has been regarded as the Satisfiability/Optimization Modulo Theories (SMT/OMT) problem, which has been addressed by calculating the static schedules with an off-the-shelf solver. Although some suitable scheduling solutions may be acquired with SMT/OMT solvers, time complexity increases rapidly if the number of flows increases [11]. Scheduling with heuristic algorithms has been discussed in literatures [12] and [13], and some solutions with relatively lower time complexity have been presented.

Since the latency of flows in TSN depends not only on the scheduling, but also on the influence from other flows to be scheduled together. To address this issue, some researchers have focused on the joint optimization of flow scheduling and routing [14]-[16]. The flow route and schedule are jointly determined, whereas the update is necessary when network structure changes or new flow occurs. Unfortunately, the delay caused by the conflicts between the time-sensitive flows in one port is rarely considered to date [17]. Moreover, current researches only focus on the schedulability analysis for simple network topology, such as automotive in-vehicle network [18] and the network with tens of flows in industrial scenarios [19].

### A. Motivation

With IIoT revolution towards next generation operational efficiency and computing connectivity, a large number of time-sensitive businesses need to be offloaded to the edge nodes for computing and analyzing. To support services with fast response time, low latency, and efficient use of bandwidth, the requirements of ultra-low latency communication in edge enabled IIoT need to be satisfied. Then how to rapidly acquire effective static schedules for massive time-sensitive flows is a major concern to be solved.

To address the above issue, in this article we analyze the scheduling problem based on the following two aspects. First, to guarantee rigorous latency requirements of time-sensitive applications, all time-sensitive and best-effort flows must be strictly separated while the performance requirements on delay and jitters must be satisfied. Existing methods designed based on SMT/OMT and Integer Linear Programming (ILP) are with high complexity due to the missing of systematical analyses on combinability for multiple flows. Then it is a great challenge to optimize the scheduling for massive time-sensitive flows in edge enabled IIoT systems. Second, mandatory decentralized reservation of bandwidth for time-sensitive flows results in the fragmentation of residual time slots, and that a static schedule for time-sensitive flows divides the hyper-frame time slot into irregular fragments would reduce the utilization of remaining bandwidth. Therefore, an appropriate scheduling algorithm is necessary to guarantee the latency of time-sensitive flows and improve the overall bandwidth utilization for edge enabled

IIoT systems.

### B. Contribution

In this article, we mainly discuss the scheduling problem of time-sensitive applications in edge enabled IIoT, and our main contributions are given as follows.

1) Theoretical analyses on the combinability problem of multiple flows are performed for the first time to our knowledge, and several conclusions about the non-collision combination of multiple time-sensitive flows are provided. Moreover, the conflicting packets between different flows can be predicted to make the scheduling scheme easy to be designed.

2) To satisfy bounded low latency and jitters requirements of time-sensitive flows, a deterministic scheduling method based on the combinability of multiple flows is presented. This method can intelligently determine whether non-conflict scheduling or the scheduling with relaxation of delay jitters is available.

3) To improve the overall bandwidth utilization, a dynamic scheduling method for non-time-sensitive flows is presented. The priorities, time-slots, and queue lengths are considered together in this method, and the port revenue is maximized to determine the most appropriate packet to be forwarded.

The remainder of this article is structured as follows. First, related works are summarized in Section II. Then we describe the problems to be solved in Section III. Section IV gives the network model and theoretical analyses. An overview and the details of the proposed scheduling methods are presented in Section V. Performance evaluations are performed for our methods in Section VI. Finally, conclusions and future works are summarized in Section VII.

## II. RELATED WORKS

Recently, research of scheduling in TSN has been a major concern due to the rapid increase of time-sensitive businesses. Existing traffic scheduling methods in TSN can be divided into two categories. One is the scheduling methods based on the Quality of Service (QoS), and another is the TAS-based scheduling methods [20]. Network calculus has been adopted

TABLE I
SUMMARY OF RELEVANT WORKS

| Analysis | Literatures | [7] | [8] | [9] | [10] | [12] | [14] | [20] | [21] | [22] | [23] | [24] | [25] |
|---|---|---|---|---|---|---|---|---|---|---|---|---|---|
| Focus | Scheduling | Yes | Yes | Yes | Yes | Yes | Yes | Yes | Yes | Yes | Yes | Yes | Yes |
| | Routing | No | Yes | No | No | No | Yes | Yes | No | Yes | Yes | No | No |
| Constraints | Constraints | No | No | No | Yes | Yes | Yes | Yes | Yes | Yes | Yes | No | No |
| Method | SMT/ OMT | No | No | Yes | Yes | Yes | No | No | Yes | No | No | No | No |
| | ILP | No | Yes | No | No | No | Yes | Yes | No | No | No | No | No |
| | Network calculus | Yes | Yes | No | No | No | No | No | No | No | No | No | No |
| Optimization | Heuristic | No | No | No | No | Yes | No | No | No | No | Yes | Yes | Yes |
| Application | Network size | Small | Small | Medium | Small | Medium | Medium | Small | Small | Small | Small | Small | Small |
| | Scenarios | Vehicle | Vehicle | IIoT | IIoT | IIoT | IIoT | Vehicle | IoT | IIoT | IIoT | IIoT | IIoT |
| Solver | Off-the-shelf | No | No | Yes | Yes | No | Yes | Yes | Yes | Yes | Yes | No | Yes |

TABLE II
APPLICATIONS REQUIREMENTS

| Applications | Latency | Jitter | Data rate | Reliability (%) |
|---|---|---|---|---|
| Motion control | <0.1ms to <2ms | <1/2 latency | 10Kbps | 99.999999 |
| Smart grid protection | 4ms to 8ms | <250us | 64Kbps | 99.9999 |
| Automated guide vehicles | 1ms to 500ms | < 1/2 latency | 10Mbps | 99.9999 |
| Safety monitor and control alarms | 5ms to 100ms | <1ms | 75Kbps | 99.99 |
| Tactile interaction | 0.5ms to 1ms | NA | 100Kbps | NA |

(NA stands for not available)

TABLE III
SOME NOTATIONS USED

| Notation | Explanation |
|---|---|
| $GCD$ | The greatest common divisor of periods for the flows |
| $T_i$ | The period of a flow $f_i$ |
| $EN_a$ | The end node $a$ in TSN |
| $SW_i$ | The $i$th switch node in TSN |
| $f_i$ | The $i$th time-sensitive flow in TSN |
| $B_i$ | The bandwidth of a flow $f_i$ |
| $B_e$ | The bandwidth of the directed edge |
| $\tau_i$ | Service time for a single packet in a flow $f_i$ |
| $o_i$ | The departure time of the 1st packet in a flow $f_i$ |
| $b_i$ | The arrival time of the 1st packet in a flow $f_i$ |
| $J$ | Delay-jitters that are accumulated before |
| $p_i$ | The packet processing time of a flow $f_i$ in the switch node |
| $F$ | Set of flows in a directed edge |
| $C$ | The delay and jitters constraints for the flows in $F$ |

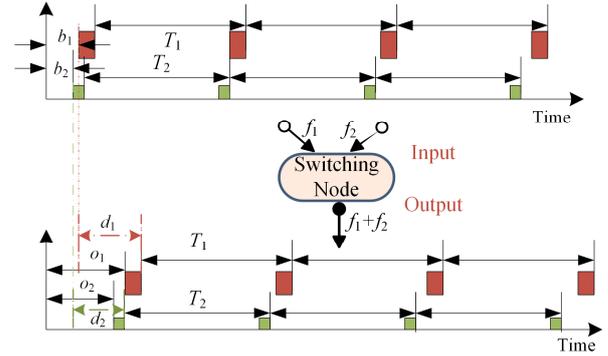

Fig. 1. An example of ideal scheduling for time-sensitive flows.

routing of flows in TSN is either considered as a programming problem with constraints or an optimization problem with the aim of minimizing the end-to-end delay of flows, as shown in Table I. Although many existing works have focused on the delay guarantee with scheduling and routing algorithms, they lack theoretic analyses on the key factors that resulting in the delay of TSN flows and do not further analyze the bandwidth utilization problem of best-effort flows. In contrast, we will theoretically analyze the combinability of time-sensitive flows to support the proposed scheduling methods, and consider the optimal combination of the strategies by using game theory to improve the overall bandwidth utilization.

## III. PROBLEM DESCRIPTION

In edge enabled IIoT systems, multiple time-sensitive flows are inevitably scheduled and forwarded through the same node port during the data routing process. To achieve deterministic forwarding, some dedicated time slots are usually preserved in advance for the time-sensitive flows, while the residual is prepared for the best-effort ones. To optimize the scheduling of flows in edge enabled IIoT, the combinability of multiple time-sensitive flows and the residual bandwidth utilization problems should be analyzed.

### A. Combinability Problem of Time-sensitive Flows

For the upcoming transformation in the context of Industry 4.0, many applications need end-to-end communications with hard-bounded guarantees for latency and reliability, as shown in Table II [4]. We call the data flows from those applications with bounded low latency and jitters as time-sensitive flows. However, fulfilling the requirement of one time-sensitive flow may make that of another flow unsatisfied. Hence, analyses on combinability of multiple flows should be performed before the design of scheduling method.

To make the analyses of the combinability problem easy to understand, we first give the definitions of some notations, as shown in Table III. The combinability of flows can directly affect the implementation of ideal scheduling. Generally, ideal deterministic scheduling for multiple time-sensitive flows in a node should satisfy two conditions: One is that the forwarding latency of a flow is as low as possible, and the other is that the original periodicity of flows is strictly maintained during the scheduling. Fig. 1 gives an example of ideal scheduling for two to analyze the worst-case latency to guarantee the QoS of time-sensitive flows in literatures [7] and [8]. The scheduling problem has been analyzed by using the SMT/OMT model while off-the-shelf solvers have been introduced to calculate static TAS schedules in Literatures [9], [10], [12], and [21].

Literature [22] has presented a method of time-sensitive software-defined network (SDN) to realize the scheduling for time-sensitive flows and best-effort flows in the same SDN. In addition, logical centralized controller has been adopted to address the global scheduling problem. ILP has been utilized to solve the routing and scheduling problems in Literature [23], it enables the terminal to execute packet scheduling of the user plane with high precision. Heuristic algorithms have been used to optimize the scheduling in literatures [12], [13], [24], and [25], with the aim of minimizing the time complexity. Joint optimization of flow scheduling and routing has been explored in Literatures [14] and [20], where the schedules of multiple flows in the routing nodes have been considered together to achieve deterministic end-to-end low latency.

Based on the reviews to existing works, the scheduling and time-sensitive flows. It shows that the periods of the flows $f_1$ and $f_2$ in an egress port keep unchanged. For ideal scheduling, each packet in a flow must be ready before the reserved time slot arrives, while the latency should be as low as possible. Moreover, the reserved time slots for packets cannot conflict with each other in the time domain.

To describe the ideal scheduling problem for the flows $f_1$ and $f_2$, an optimization objective function can be defined as

$$[o_1, o_2] = \arg\min_{o_1, o_2}(o_1 - b_1, o_2 - b_2)$$

$$s.t. \begin{cases} C_1 \text{ or } C_2 \text{ and } C_3 \\ C_1 : \tau_2 < (nT_1 + o_1) - (mT_2 + o_2) \\ C_2 : \tau_1 < (mT_2 + o_2) - (nT_1 + o_1), \quad \forall n, m \in \mathbb{N}, \\ C_3 : \begin{cases} o_1 - b_1 > J_1 + p_1 \\ o_2 - b_2 > J_2 + p_2 \end{cases} \end{cases} \quad (1)$$

where $o_1$ and $o_2$ denote the departure time of the 1st packets in the flows $f_1$ and $f_2$, respectively; $b_1$ and $b_2$ are the arrival time of the 1st packets; $T_1$ and $T_2$ are periods of $f_1$ and $f_2$, respectively; $J_1$ and $J_2$ are cumulated jitters before; $p_1$ and $p_2$ are the packet processing time of $f_1$ and $f_2$, respectively; $\tau_1$ and $\tau_2$ are the packet service time for $f_1$ and $f_2$ in the egress port, respectively. Here it is assumed that the minimum time unit is 1, and all time variables are integers.

Ideal scheduling can effectively avoid delay jitters, because the $n$th packet in the flow $f_i$ is forwarded within the time slot $[nT_i + o_i, nT_i + o_i + \tau_i]$ so that all packets in the same flow have the same latency. However, whether multiple flows can be ideally scheduled together needs to be further analyzed since the conflicts between two flows may exist. The formal definition of the conflict is given as follows:

***Definition 1:*** Conflict between two flows $f_1$ and $f_2$ occurs if there exist two positive integers $(m, n)$ that make either of the following two conditions true.

1) $nT_1 + o_1 = mT_2 + o_2$,

2) $0 < (nT_1 + o_1) - (mT_2 + o_2) < \tau_2$ or
   $0 < (mT_2 + o_2) - (nT_1 + o_1) < \tau_1$.

If the first condition is true, conflict of the first kind (CFK) occurs. If and only if the second is true, conflict of the second kind (CSK) appears. If neither CFK nor CSK occurs, multiple flows can be scheduled together to achieve ideal scheduling.

### B. Residual Bandwidth Utilization (RBU) Problem

To guarantee deterministic delay, time slots for all packets in each time-sensitive flow should first be assigned. Then the residual time slots can be shared by other best-effort flows. Whereas, this mandatorily allocation method inevitably results in the fragmentation of the residual time slots.

For traditional scheduling methods like Strict Priority (SP) and Round Robbin (RR), the remaining bandwidth is the only metric to measure whether flows are schedulable or not since it is continuously concentrated in the time domain. While the remaining bandwidth for deterministic scheduling is usually dispersed in the time domain. If the length of the packet to be scheduled does not match the size of a residual time slot, then this time slot may be wasted or partially occupied. Hence, to improve the remaining bandwidth utilization, it is necessary to take features of potential best-effort flows into consideration.

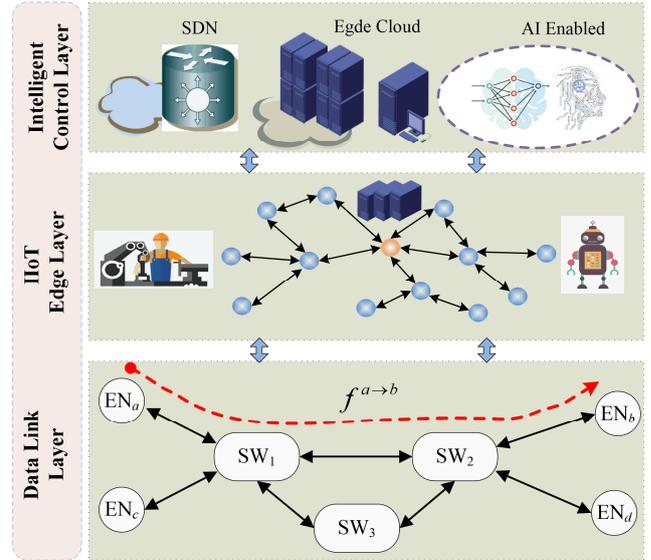

Fig. 2. An edge enabled IIoT framework in time-sensitive scenarios

## IV. NETWORK MODEL AND THEORETICAL ANALYSES

In edge enabled IIoT, various time-sensitive applications such as multi-agent communication, key-value collection, and remote real-time controlling increase rapidly. An edge enabled IIoT framework in time-sensitive scenarios is given in Fig. 2. For any time-sensitive application, the intelligent control layer first provides the solutions for the centralized configuring, intelligent scheduling, and SDN based routing according to the corresponding requirements. Then communication services are provided in edge layer [26], [27]. The data link layer finally performs packet delivering to guarantee latency requirements of the time-sensitive flows.

### A. Network Model

A typical TSN network can be considered as the scheduling platform of time-sensitive applications, and it can be modeled as a graph $G = \{V, E\}$, where $V$ is the vertex set consisting of end nodes and switch nodes, and $E$ is the set of directed edges [10], [28]. Each directed edge denotes a physical link that connects two vertices.

An end-to-end unicast or multicast network flow from one end node $EN_a$ to another $EN_b$ via $n$ switch nodes ($SW_1 \dots SW_n$), expressed as $f^{a \to b}$, can be considered as a tuple. A tuple $f^{a \to b}$ is denoted by a set $\{f^{a \to b}.R, f^{a \to b}.\varphi, f^{a \to b}.P\}$, where $f^{a \to b}.R$ is the directed subgraph $\{EN_a \to SW_1 \to SW_2 \to \dots \to SW_n \to EN_b\}$ that describes the transmission path from $EN_a$ to $EN_b$ via $n$ switch nodes; $f^{a \to b}.\varphi$ is the set of emerged time offsets when a flow passes through the egress ports of all switch nodes; $f^{a \to b}.P$ is the set of all packets in the flow. For an intermediate switch node $SW_m$, the set of input flows is expressed as $\{f.\rightarrow SW_m\}$, and the set of output flows is $\{SW_m f.\rightarrow\}$.

### B. Theoretical Analyses on Combinability of Multiple Flows

The combinability means no conflict occurs for the flows transmitted via the same directed edge. While the conflict may be inevitable for the time-sensitive flows. To address this issue, we give some theorems below.

***Theorem 1:*** For the flows $f_1$ and $f_2$, if their periods $T_1$ and $T_2$ are co-prime ($GCD(T_1, T_2) = 1$), the conflict CFK occurs, and if $\tau_1 > 1$ or $\tau_2 > 1$, the conflict CSK occurs.

***Proof:*** For any two flows $f_1$ and $f_2$, if $GCD(T_1, T_2) = 1$, then according to the Bezout's Theorem [29], $\exists n, m \in N$ make $nT_1 + o_1 = mT_2 + o_2$ true, therefore, CFK occurs. If $GCD(T_1, T_2) = 1$ and $\tau_1 > 1$, then $\exists n, m \in N$ make $(mT_2 + o_2) - (nT_1 + o_1) = 1$ true. That is, $0 < (mT_2 + o_2) - (nT_1 + o_1) < \tau_1$ is true and CSK occurs.

***Theorem 2:*** For the flows $f_1$ and $f_2$, if CFK occurs, then the number pairs of the packets with CFK can be expressed as a set $\{(n, m) | n = n_0 + kT_2/GCD(T_1, T_2), m = m_0 + kT_1/GCD(T_1, T_2)\}$, $k \in N$. Here $(n_0, m_0)$ is a particular number pair of the packets with CFK from $f_1$ and $f_2$.

***Proof:*** Given two flows $f_1$ and $f_2$, if CFK occurs, $\exists n_0, m_0 \in N$ make the Bezout's Equation [30] $nT_1 + o_1 = mT_2 + o_2$ true. The complete solution of the equation can be achieved as $[n, m] = [n_0 + kT_2/GCD(T_1, T_2), m_0 + kT_1/GCD(T_1, T_2)]$.

***Lemma 1:*** Non-collision combination of two time-sensitive flows is available if all the following conditions are satisfied:

1) $g = GCD(T_1, T_2) > 1$,

2) $\begin{cases} \tau_2 \le (o_1 - o_2)\% g \le g - \tau_1, & o_1 > o_2 \\ \tau_1 \le (o_2 - o_1)\% g \le g - \tau_2, & o_1 < o_2 \end{cases}$,

3) $\tau_1 + \tau_2 < g$.

***Proof:*** We try to prove the above lemma with the method of reduction to absurdity. Assuming that all the conditions above are satisfied for two flows with CFK or CSK. If CFK occurs, then $\exists n, m \in N$ make $nT_1 + o_1 = mT_2 + o_2$ true. Hence $nT_1 - mT_2 = o_2 - o_1$, and $nT_1 - mT_2 = kg$, so that $(o_1 - o_2)\% g = 0$. That is, if CFK occurs, the second condition is not satisfied. Similarly, it can be proved that if CSK occurs, the second condition cannot be satisfied. Therefore, the assumption is invalid.

***Theorem 3:*** Non-collision combination of $K$ time-sensitive flows is available if all the following conditions are satisfied:

1) $g = GCD(T_1, \cdots, T_k) > 1$,

2) $\forall n, m = 1, \dots, K$
$\begin{cases} \tau_m \le (o_n - o_m)\% g \le g - \tau_n, & o_n > o_m \\ \tau_n \le (o_m - o_n)\% g \le g - \tau_m, & o_n < o_m \end{cases}$,

3) $\sum_{i=1}^{K} \tau_i < g$.

The proof of ***Theorem 3*** is similar to ***Lemma 1***.

If both the first and the third conditions in ***Theorem 3*** are satisfied, then there must exist a set $\{o_i\}$ that makes the second condition be satisfied.

***Theorem 4:*** For $K$ time-sensitive flows, if $GCD(T_1,...,T_K) = 1$, CFK occurs, and if $\exists \tau_i > 1$ $(i=1,..., K)$, CSK occurs.

The proof of ***Theorem 4*** is similar to ***Theorem 1***.

The worst combination of $K$ time-sensitive flows is that the periods of any two flows are co-prime so that CFK will occur for the $K$ flows. It can be described as a linear equation set that has integer solution space. Therefore, we have

$$\begin{cases} T_1 x_1 - T_2 x_2 = o_2 - o_1 \\ T_2 x_2 - T_3 x_3 = o_3 - o_2 \\ \quad \vdots \\ T_{K-1} x_{K-1} - T_K x_K = o_K - o_{K-1}, \end{cases} \tag{2}$$

and (2) can be rewritten as

$$AX = b, \tag{3}$$

where

$$A = \begin{bmatrix} T_1 & -T_2 & 0 & \cdots & 0 & 0 \\ 0 & T_2 & -T_3 & 0 & \cdots & 0 \\ \vdots & \vdots & \ddots & \ddots & \vdots & \vdots \\ 0 & 0 & \cdots & T_{K-2} & -T_{K-1} & 0 \\ 0 & 0 & \cdots & 0 & T_{K-1} & -T_K \end{bmatrix},$$

$$b = \begin{bmatrix} o_2 - o_1, o_3 - o_2, \cdots, o_{K-1} - o_{K-2}, o_K - o_{K-1} \end{bmatrix}^T.$$

The integer solution space of (3) consists of homogeneous general solution and non-homogeneous particular solution, and it can be expressed as

$$X = k \begin{bmatrix} a_1 \\ a_2 \\ \vdots \\ a_{K-1} \\ a_K \end{bmatrix} + \begin{bmatrix} x_1^0 \\ x_2^0 \\ \vdots \\ x_{K-1}^0 \\ x_K^0 \end{bmatrix}, k \in N. \tag{4}$$

A recursive method is adopted to solve (3). For the case that CFK between two flows ($K=2$) occurs, the integer solution can be obtained based on ***Theorem 2***, which is of the form

$$X = k_2 \begin{bmatrix} T_2 \\ T_1 \end{bmatrix} + \begin{bmatrix} x_1^0 \\ x_2^0 \end{bmatrix}, k_2 \in N. \tag{5}$$

The form of integer solution space of $n$-1 linear equations can be acquired based on (4). To search the solution of $n$ linear equations, another equation is added that expressed as follows:

$$T_n x_n - T_{n+1} x_{n+1} = o_{n+1} - o_n. \tag{6}$$

The solution of (6) is

$$\begin{bmatrix} x_n \\ x_{n+1} \end{bmatrix} = k_{n+1} \begin{bmatrix} T_{n+1} \\ T_n \end{bmatrix} + \begin{bmatrix} x_n^{p0} \\ x_{n+1}^{p0} \end{bmatrix}. \quad (7)$$

We can obtain $x_n$ according to (4)

$$x_n = k a_n + x_n^0. \quad (8)$$

Then we can easily get the solutions to $k$ and $k_{n+1}$ that can make both (7) and (8) be satisfied:

$$\begin{bmatrix} k \\ k_{n+1} \end{bmatrix} = k' \begin{bmatrix} T_{n+1} / GCD(a_n, T_{n+1}) \\ a_n / GCD(a_n, T_{n+1}) \end{bmatrix} + \begin{bmatrix} t_n^0 \\ t_{n+1}^0 \end{bmatrix}, \quad (9)$$

where $k' \in N$, $t_n^0$ and $t_{n+1}^0$ are non-homogeneous particular solutions to the coefficient variables $k$ and $k_{n+1}$.

That $a_n$ and $T_{n+1}$ are co-prime can also be proved using the recursive method. Then (7) can be rewritten as

$$\begin{bmatrix} x_n \\ x_{n+1} \end{bmatrix} = k' \begin{bmatrix} T_{n+1} a_n \\ T_n a_n \end{bmatrix} + \begin{bmatrix} a_n t_n^0 + x_n^0 \\ T_n t_{n+1}^0 + x_{n+1}^{p0} \end{bmatrix}. \quad (10)$$

Finally, we can get the integer solution space of (3):

$$X = k \begin{bmatrix} \prod_{i \neq 1} T_i \\ \prod_{i \neq 2} T_i \\ \vdots \\ \prod_{i \neq K-1} T_i \\ \prod_{i \neq K} T_i \end{bmatrix} + \begin{bmatrix} \sum_{m=2}^{K-1} (\prod_{j \neq 1}^{j=m+1} T_j) t_{1m}^0 + x_1^{p0} \\ \sum_{m=2}^{K-1} (\prod_{j \neq 2}^{j=m+1} T_j) t_{2m}^0 + x_2^{p0} \\ \vdots \\ \sum_{m=2}^{K-1} (\prod_{j \neq K-1}^{j=m+1} T_j) t_{(K-1)m}^0 + x_{K-1}^{p0} \\ \sum_{m=2}^{K-1} (\prod_{j \neq K}^{j=m+1} T_j) t_{Km}^0 + x_K^{p0} \end{bmatrix}, \quad (11)$$

where $x_i^{p0}$ and $t_{im}^0$ $(i = 1, 2, ..., K)$ are the particular solutions of the $i$th equation in (2) and the related coefficient variable $k_m$ in the $m$th recursive calculation, respectively.

The worst combination of $K$ time-sensitive flows is that the periods of any two flows are co-prime, and the CFK among $K$ flows occurs every $\prod T_i$.

Similarly, if there exists CSK among $K$ flows, the following linear equation (12) has integer solution space.

$$\begin{cases} T_1 x_1 - T_2 x_2 = o_2 - o_1 + v_1 \\ T_2 x_2 - T_3 x_3 = o_3 - o_2 + v_2 \\ \vdots \\ T_{K-1} x_{K-1} - T_K x_K = o_K - o_{K-1} + v_{K-1}, \end{cases} \quad (12)$$

where $|v_i|$ $(i = 1, ..., K-1)$ denotes the overlap length in time domain. It should be noted that, here $|v_i|$ should be non-zero, if not, it means that CFK exists. We can get the solution space of (12) using the similar method adopted for (2).

TABLE IV
FLOW PARAMETERS

| Flow $f_i$ | $T_i$ | $\tau_i$ | $o_i$ |
|---|---|---|---|
| $f_1$ | 14 | 3 | 0 |
| $f_2$ | 27 | 3 | 5 |
| $f_3$ | 61 | 4 | 9 |

To make it easy to understand, an example is given here to explain how to get the solution. Assuming that there are three flows to be scheduled, and their periods are co-prime so that CFK exists. The flow parameters are shown in Table IV. We can get the linear equations about the worst combination:

$$\begin{cases} 14 x_1 - 27 x_2 = 5 - 0 \\ 27 x_2 - 61 x_3 = 9 - 5 \end{cases}. \quad (13)$$

The integer solution of $14 x_1 - 27 x_2 = 5$ is as follows:

$$\begin{bmatrix} x_1 \\ x_2 \end{bmatrix} = k_1 \begin{bmatrix} 27 \\ 14 \end{bmatrix} + \begin{bmatrix} 10 \\ 5 \end{bmatrix}, k_1 \in N. \quad (14)$$

Similarly, we can get the integer solution of $27 x_2 - 61 x_3 = 4$ as follows:

$$\begin{bmatrix} x_2 \\ x_3 \end{bmatrix} = k_2 \begin{bmatrix} 61 \\ 27 \end{bmatrix} + \begin{bmatrix} 25 \\ 11 \end{bmatrix}, k_2 \in N. \quad (15)$$

Then we can easily get the solutions to $k_1$ and $k_2$ that can make (14) match (15):

$$\begin{bmatrix} k_1 \\ k_2 \end{bmatrix} = k \begin{bmatrix} 61 \\ 14 \end{bmatrix} + \begin{bmatrix} 45 \\ 10 \end{bmatrix}, k = 1, 2, 3, \cdots. \quad (16)$$

Therefore, the solution space of (13) is

$$\begin{bmatrix} x_1 \\ x_2 \\ x_3 \end{bmatrix} = k \begin{bmatrix} 27*61 \\ 14*61 \\ 14*27 \end{bmatrix} + \begin{bmatrix} 27*45+10 \\ 14*45+5 \\ 27*10+11 \end{bmatrix}, k = 0, 1, 2, \cdots. \quad (17)$$

Based on the results, we can confirm the conflicting packets. For example, if $k = 0$, CFK occurs for the packets numbered [27*45+10, 14*45+5, 27*10+11] from the three flows.

To the best of our knowledge, this is the first time to make theoretical analyses for combinability of $K$ time-sensitive flows, which will guide the scheduling design and provide theoretical basis. The conditions of non-collision combination for flows are provided in **Theorem 3**. For multiple flows to be scheduled together, we first consider whether these conditions are true. If they are, jitters-free scheduling may be available for the flows. Otherwise, that all conflicting packets can be confirmed in advance makes it easy to eliminate future conflicts during the scheduling.

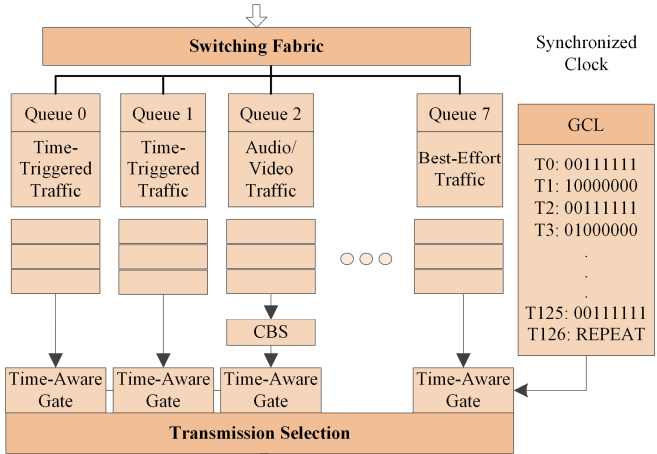

Fig. 3. Egress port with IEEE 802.1Qbv enhancements.

## V. PROPOSED SCHEDULING METHOD

### A. Data Forwarding Model

Before giving the scheduling model, let us present some hypotheses and preconditions. First, we assume that all the switch nodes are rational and selfish individuals with the aim of maximizing the revenue. Second, the benefit of transmitting a time-sensitive packet in the reserved time slot is much higher than that of transmitting a best-effort packet. Therefore, the time-sensitive flows should not be disturbed by the best-effort ones. Third, if any packet is dropped in a port, the punishment should be imposed on the corresponding switch node.

Deterministic delay services usually coexist with best-effort ones. To assign the two kinds of services into a schedule, a dynamic scheduling method needs to be presented. We give a data forwarding model for switch nodes to make decision. A game model $G = <P, S, U>$ is formally defined. Where $P$ is the payers consisting of $n$ players in an egress port; $S = \{s_i\}$ is the strategy set; and $U = \{u_i\}$ is the utility function set. It is assumed that, as the player, each port is a rational and selfish intelligent individual [31]. Then each player always expects to improve its own payoff, which not only depends on its own strategy, but also subjects to the states of others.

For an egress port, a packet from one queue will be chosen for transfer if it is ready while the port is free. The strategy determined by the queue is a finite pure strategy that is either "ready for transfer" or "not ready". If a packet in a queue $i$ is selected to be ready for transfer, then the strategy is $s_i = 1$; otherwise, $s_i = 0$. The strategy subjects to $\sum s_i \le 1$, $\forall s_i = 0$ or 1.

Each queue in an egress port ends with a time-aware gate controlled by Gate Control List (GCL) in TSN [26], as shown in Fig. 3. Every row in GCL contains some gate sates that are either open or close [8]. From the queues whose gates are open, packets are selected sequentially and ready for transfer based on the designed scheduling method. We try to give a dynamic transmission selection algorithm to keep a better coexistence for the time-sensitive and best-effort flows.

### B. Non-collision Deterministic Scheduling (NDS)

Time-sensitive flows always have the strict delay and jitters constraints. Then deterministic scheduling methods need to be designed for the purpose of providing services with bounded low latency and low jitters for the flows. The assignment of time slots for time-sensitive flows is crucial to satisfy the delay requirements. To address this issue, we present a non-collision deterministic scheduling (NDS) method for TSN flows.

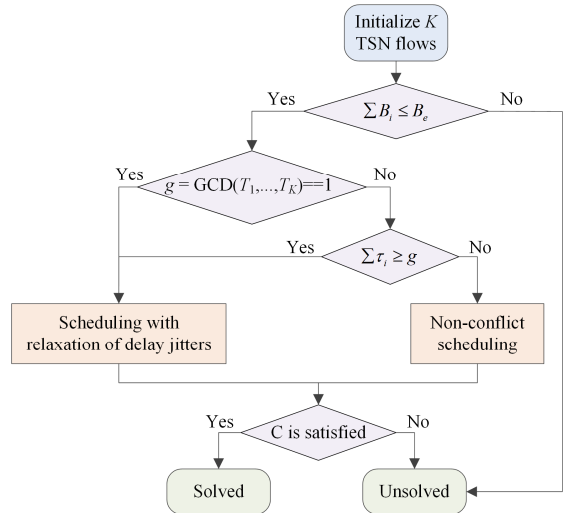

Fig. 4. Overview of the proposed NDS method.

An overview of the proposed NDS method is given in Fig. 4. Since bandwidth constraint must be satisfied for deterministic scheduling, the total bandwidth occupied by all time-sensitive flows cannot exceed the bandwidth of a directed edge. If the greatest common divisor $g$ is greater than 1, while the total service time for $K$ packets, which are separately selected from $K$ time-sensitive flows, is less than $g$, then non-conflict ideal scheduling method can be considered. Otherwise, scheduling method with relaxation of delay jitters is adopted.

For a given flow set $F$ that contains $K$ time-sensitive flows, the static schedule calculation method is given in Algorithm 1. If conditions 1 and 3 in **Theorem 3** are satisfied, non-conflict ideal scheduling method can be adopted. Then the time-offset set $O$ needs to be determined using the presented method in Algorithm 2. Based on the acquired set $O$, the static schedules can be directly computed. For example, when the $n$th packet of the flow $f_i$ in a queue is ready for transfer, the gate of the queue will be open within the time slot $[nT_i + o_i, nT_i + o_i + \tau_i]$. Otherwise, scheduling method with relaxation of delay jitters has to be adopted. First, the flow set $F$ needs to be divided into several subsets. Then, to restrict the occurrence of CFK and CSK, non-conflict ideal scheduling is used for every subset if all applicable conditions are true. Next, to eliminate conflicts between flows, a searching method is adopted to perform relaxation operation of the constrained delay jitters. The final static schedule within a cycle that is the least common multiple of periods for the flows in set $F$ is determined.

For a flow set $F$, it needs to be divided into several subsets if either CFK or CSK is inevitable. The flows that satisfy the combinability can be grouped into the same subset. Moreover, each subset from $F$ expects to contain as many combinative flows as possible. Then the time complexity of the scheduling

---

**Algorithm 1: Static Schedule Calculation for K flows.**

---

**Input:** Set $F$ of $K$ flows, initial time-offset set $\varphi$, delay-jitters constraint set $C$, processing time set $P$
**Output:** The static schedule $SCH$
1: **if** $GCD(F) > 1 \wedge \sum \tau_j \leq g$ **then**
2:     $(O, S) \leftarrow$ Non-conflict schedule $(F, \varphi, C, P)$;
3:     $SCH \leftarrow \{[nT_i + o_i, nT_i + o_i + \tau_i]\}$;
4: **else**
5:     $\{F_1, F_2, ..., F_m\} \leftarrow$ set-partition$(F)$;
6:     $\{g_1, g_2, ..., g_m\} \leftarrow \{GCD(F_i) \mid i = 1, 2, ..., m\}$;
7:     Initialize $O$;
8:     **for-each** $F_i$ (from 1 *to* $m$)
9:         **if** $g_i > 1 \wedge \sum \tau_j \leq g_i$ **then** in remaining time slots,
10:             $(O_i, S) \leftarrow$ Non-conflict schedule $(F_i, \varphi_i, C_i, P_i)$;
11:         **end if**
12:     **end for-each**
13:     **Loop** to search $SCH$ until $C$ is satisfied or time out for all packets within the least common multiple ($LCM$) of periods for flows in $F$
14:         **if** CFK or CSK occurs **then**
15:             Eliminate the conflicts based on $C$;
16:         **end if**
17:     **end Loop**
18: **end if**
19: **if** $S == 0$ or $C$ is satisfied **then**
20:     Output the $SCH$;
21: **end if**

---

**Algorithm 2: Non-Conflict Time-Offset Set Calculation.**

---

**Input:** $F_i, \varphi_i, C_i, P_i$
**Output:** Time-offset set $O$ of $F$, unsolved identifier $S$
1: $g \leftarrow GCD(T_1, T_2, ..., T_n)$;
2: $l \leftarrow LCM(T_1, T_2, ..., T_n)$;
3: $o_i \leftarrow \varphi_i + p_i$;
4: **for-each** $o_i \in O$
5:     **while** $(o_i < T_i)$
6:         **if** for all $o_j (j \neq i), (k-1)g + \tau_i \leq o_i - o_j \leq kg - \tau_j$
7:             stop;
8:         **else**
9:             $o_i$++;
10:            **if** $o_i == c_i$ **then**
11:                $S = 1$;
12:            **end if**
13:        **end if**
14:    **end while**
15: **end for-each**
16: **if** $S != 1$ **then**
17:     $O \leftarrow \{o_1, o_2, ..., o_K\}$;
18: **end if**
19: **return** $O, S$;

---

can be further reduced while the requirements for delay-jitters are easier to be satisfied.

### C. Dynamic Queue Scheduling (DQS)

The above method NDS aims at guaranteeing the latency of time-sensitive flows through analyzing the combinability of multiple flows. Whereas, mandatorily allocation of bandwidth for time-sensitive flows inevitably results in the fragmentation of the residual time slots. To improve the overall bandwidth utilization, other best-effort flows with QoS requirements need to be elaborately scheduled within the residual time slots. To address this issue, we present a dynamic queue scheduling (DQS) method for those non-time-sensitive flows.

For the scheduling of non-time-sensitive flows, the strategy of an egress port not only depends on the present available time slots and the number of packets that are ready for transfer in the queues, but also is related to the strategies of the ports in neighbor nodes. In addition, the number of packets that are ready for transfer in a queue may be affected by the present strategy and the number of packets arrived from the neighbors in the followed period. To address this issue, reinforcement learning is borrowed to design the dynamic queue scheduling.

For scheduling, a port always expects to maximize its own payoff that may be affected by the present and next strategies. Therefore, the present and next strategies should be considered when designing utility functions. The present utility function for the strategy $s$ is first given as follows:

$$u^{pr} = c \cdot s^T - \beta p \cdot \left( ones(1, n) - s^T \right), \qquad (18)$$

where $c = [c_1 c_2 \ldots c_n]$ is the beneficial coefficient vector, and $p = [p^0 p^1 \ldots p^n]$ is the penalty factor vector for packet loss.

Penalty factor vector $p$ depends on the number of packets in queues, then penalty factor $p^i$ regarding queue $i$ is of the form

$$p^i = \begin{cases} 0 & q_L^{max}/2 \geq q_L^i \\ q_L^i / q_L^{max} & q_L^{max}/2 \leq q_L^i < q_L^{max} \\ p_0 & q_L^{max} \leq q_L^i \end{cases}, \qquad (19)$$

where $q_L^i$ is the number of packets ready for transfer, $q_L^{max}$ and $p_0$ are the maximum queue length and penalty, respectively.

The utility function of next strategy $\bar{s}$ can be expressed as

$$u^{ns} = c \cdot s^T - \beta (\bar{p} - s) \cdot \left( ones(1, n) - \bar{s}^T \right), \quad (20)$$

where $\bar{p}$ is the predicted penalty factor vector, and the $i$th factor is of the form

$$\bar{p}^i = \begin{cases} 0 & \dfrac{q_L^{max}}{2} \geq q_L^i + \tau E_a^T \\ \dfrac{(q_L^i + \tau E_a^i(t))}{q_L^{max}} & \dfrac{q_L^{max}}{2} \leq q_L^i + \tau E_a^T < q_L^{max} \\ p_0 & q_L^{max} \leq q_L^i + \tau E_a^T \end{cases}, \quad (21)$$

where $E_a(t)$ is the expected number of the arrival packets in time slot $t$, and $E_a^T$ is the expected number within the period $T$.

The final utility function of the port is a combination of the above two utility functions, which can be expressed as

$$u = \alpha u^{pr} + (1-\alpha) u^{ns}. \qquad (22)$$

Hence, the best strategy $s*$ of a port can be obtained by

TABLE V
PARAMETERS SETTING

| Flows number | Periods $T$ | Ratios | Packet sizes |
|---|---|---|---|
| 5 | (0.5ms, 2ms, and 5ms) | 2:2:1 | 64 to 512B |
| 20 | (0.5ms, 2ms, and 5ms) | 6:7:7 | 64 to 512B |
| 50 | (0.5ms, 2ms, and 5ms) | 16:17:17 | 64 to 512B |
| 100 | (0.5ms, 2ms, and 5ms) | 33:33:34 | 64 to 512B |

maximizing its utility:

$$s^* = \underset{s}{arg\ max}(u). \tag{23}$$

For our scheduling methods NDS/DQS, the deterministic delay of time-sensitive flows can be guaranteed using the NDS method. Lower priorities are assigned to the best-effort flows, which are scheduled in the remaining decentralized time slots. Both the fragmentation of the remaining time slots and the dynamic characteristic regarding the number of ready packets in queues are considered in the DQS method. Services for the best-effort flows are expected to be guaranteed while the utilization of the remaining bandwidth can be improved.

## VI. SIMULATION RESULTS AND ANALYSIS

In this section, we perform the experiments to evaluate the performance of our scheduling algorithms. Experiments run on an Intel(R) Core(TM) i5-2600 64bit CPU@3.4GHz with 12GB of RAM via MATLAB R2020b.

### A. Simulation Setup

In the simulations, we evaluate the delay and jitters of time-sensitive flows and the bandwidth utilization of best effort flows in a switch node. All egress ports are full-duplex and with 1Gbit/s bandwidth.

We refer to the descriptions of flow properties in IEC/IEEE 60802 to simulate the flows in actual industrial scenarios. For time-sensitive flows, the packet sizes range from 64 to 512B, the periods are 0.5ms, 2ms or 5ms, the arrival time of packets is randomly generated. To simulate the flows under different industrial scenarios, we count the delay and jitters for 5 to100 flows on a single directed edge, as shown in Table V, and the ratio of flows with different periods is nearly 1:1:1.

### B. Performance of NDS Method

To better evaluate the proposed flow scheduling algorithms, we focus on the forwarding delay for a single switch node and the flow scheduling on a single directed edge. Assuming that there are $m$ flows to be forwarded in a switch node $SW_i$ and $n$ flows need to be scheduled on the directed edge $SW_i \rightarrow SW_j$.

The results of delay for all flows scheduled on $SW_i \rightarrow SW_j$ using the proposed NDS, D/MM [11], DA/IRS [14], and EPIC [28] algorithms are shown in Fig. 5. To verify the performance of our proposed algorithm, we count the maximum, minimum, and average delay of 20 time-sensitive flows assigned to one directed edge. We use error bar to show deterministic delay of flows for D/MM and DA/IRS. These flows are approximately divided into three even groups with the periods 0.5ms, 2ms, and 5ms. The packet sizes of these flows range from 64 to 512. Compared with the other three algorithms, NDS can provide Non-conflict and jitters-free scheduling for flows as expected. Because the total service time for the 20 packets, which are separately selected from the above 20 time-sensitive flows, is less than the greatest common divisor of periods (0.5ms, 2ms, and 5ms), then the first and third conditions in **Theorem 3** are satisfied. Hence, it is certain that a time-offset set $\{o_i\}$ exists to make the second condition satisfied. The forwarding delay of each packet from the 20 flows is counted in the experiment. The results for the maximum, minimum, and average delay of every flow are shown in this figure. The EPIC algorithm can also achieve the jitters-free scheduling for flows. However, the delay using EPIC is much higher than NDS. Jitters of the 20 flows using D/MM and DA/IRS are relatively higher than the average delay. This indicates that the two algorithms have a weaker performance in guaranteeing deterministic forwarding services than NDS and EPIC, since the jitter of a flow is the difference between the maximum and minimum delay. The delay using NDS is lower than the delay using EPIC and the average delay using D/MM and DA/IRS, this means that the non-collision combination theory can better guide scheduling optimization.

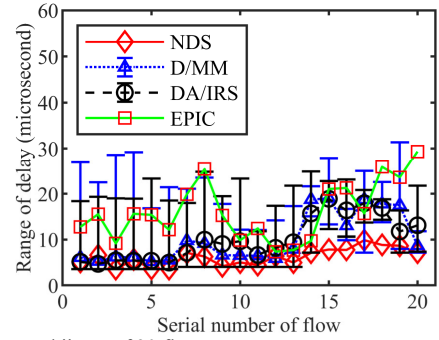

Fig. 5. Delay and jitters of 20 flows.

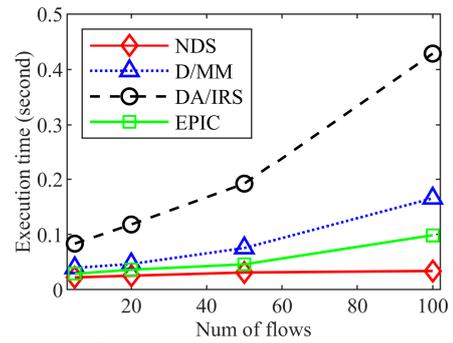

Fig. 6. Comparison of the execution time.

Taking the machine control as an example, the commands should be transmitted to the destination within 0.5ms while jitters should be avoided. Fig. 5 shows jitters-free scheduling can be achieved using the proposed algorithm NDS, and the delay is no more than 15us for 20 flows scheduled together in a node. Therefore, algorithm NDS can meet the requirements of the time-sensitive applications in intelligent manufacturing, which is essential for edge enabled IIoT.

The comparison of execution time for these algorithms is given in Fig. 6. Here the number of time-sensitive flows ranges from 5 to 100. It can be noticed that, with the increase of the number of flows, the execution time of D/MM, DA/IRS, and EPIC increases faster than that of NDS. The results also show that the average execution time of NDS for 100 flows is 33.7ms, while those of EPIC, D/MM, and DA/IRS are 98.6ms, 165.5ms, and 428.6ms, respectively. This big gap owes to that D/MM and DA/IRS search the qualified solutions for all packets in a superperiod (the least common multiple of periods for total 100 flows) without/with delay and jitters constraints, and EPIC searches the solutions by seeking for time offsets to make any two flows non-collision. While NDS only computes these appropriate time offsets of flows so that the complexity is greatly simplified and the execution time can be reduced.

The comparison of schedulable ratios among NDS, D/MM, DA/IRS, and EPIC is shown in Fig. 7. It shows that, the schedulable ratio of NDS keeps high while those of D/MM, DA/IRS, and EPIC decrease with the increase of the number of flows. In our proposed algorithm NDS, the combinability of multiple flows is analyzed. When all conditions in **Theorem 3** are satisfied, the schedule can be computed only by searching for the optimal time-offset set. Otherwise, relaxation of jitters needs to be considered to resolve the conflicts. Then packets with CFK or CSK can be confirmed via theoretical deduction, so that the optimal solutions of scheduling are easy to search. This figure also shows that D/MM has the lowest schedulable ratio for all cases since it ignores delay and jitters constraints, which are considered by all other algorithms.

### C. Performance of DQS Method

To improve the overall bandwidth utilization, a scheduling method DQS for the best-effort flows is also presented in this article. In our proposed method, the present and next strategies are considered together to maximize the bandwidth utilization. The weight coefficients of the benefits for the present and next strategies are both set as 0.5. To evaluate the performance, experiment is performed under the case that the number of time-sensitive flows ranges from 5 to 100 while the percent of bandwidth required for the best-effort flows is nearly 50. Note that, the bandwidth of the directed edge is enough for all time-sensitive and best-effort flows. Since priority-based queue forwarding mechanism is adopted in SP, the time slot can be fully occupied if there are enough packets ready for transfer. Then SP is considered as a benchmark for the comparison of bandwidth utilization among these algorithms.

The comparison of bandwidth utilization is given in Fig. 8. It shows that SP has the highest bandwidth utilization, while the gap of bandwidth utilization between SP and the other three algorithms increases with the increase of the number of time-sensitive flows. This figure also shows that NDS/DQS has higher bandwidth utilization than D/MM, DA/IRS, and EPIC. This is because the present and next strategies are jointly considered in NDS/DQS to reduce the number of idle time slots. To some extent, the proposed method can reduce the ratio of packet loss caused by queuing, especially for the best-effort flows with low priorities and long packets. Hence, the proposed method NDS/DQS can help to realize ultra-low latency communication in TSN networks.

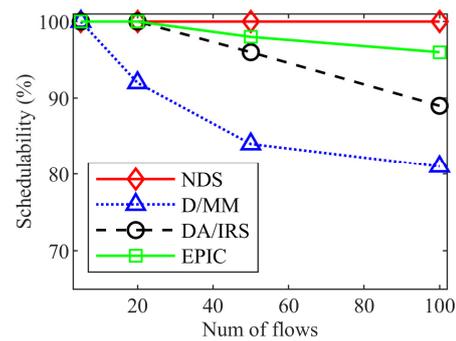

Fig. 7. Comparison of schedulable ratios.

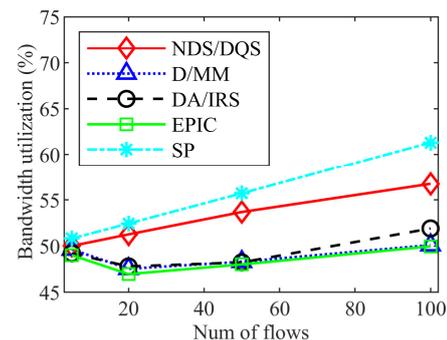

Fig. 8. Comparison of bandwidth utilization.

## VII. CONCLUSION

To support the services with fast response time, low latency, and efficient use of bandwidth in edge enabled IIoT, in this article, we present the traffic scheduling method NDS/DQS to guarantee the ultra-low latency communication. The proposed method includes deterministic scheduling algorithm NDS for the time-sensitive flows and queue scheduling algorithm DQS for the best-effort flows. The combinability of time-sensitive flows is analyzed and formalized in NDS while the RBU problem is concerned in DQS. Experiments are performed to verify the performances regarding delay and jitters, execution time, schedulable ratio, and bandwidth utilization. The results demonstrate the effectiveness and outperformance of the proposed method in scheduling for both time-sensitive and best-effort flows.

In future work, we will further research ultra-low latency communication and study the deterministic latency guarantee techniques for IIoT. The applications of AI in scheduling will be explored to optimize the delay and jitters for time-sensitive flows. To ensure the end-to-end latency, deterministic routing protocols based on the hypergraph model will be our research focus.

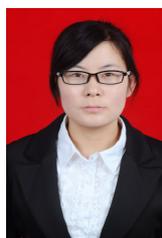

**Yinzhi Lu** received her M.S. degree in Communication and Information Systems from Chongqing University, Chongqing, China, in 2014. She is currently pursuing the Ph. D. degree in Information and Communication Engineering with the School of Communication and Information Engineering, Chongqing University of Posts and Telecommunications. She was a teaching assistant with the School of Electronic Information Engineering, Yangtze Normal University from 2014 to 2019. Her current research interests include Internet of Things, time sensitive network, and artificial intelligence.

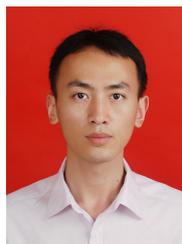

**Liu Yang** received his B.S. degree in Electronic Information Science and Technology from Qingdao University of Technology, Shandong, China, in 2010, and Ph.D. degree in Communication and Information Systems at the School of Communication Engineering, Chongqing University, Chongqing, China, in 2016. He is now a lecturer in Chongqing University of Posts and Telecommunications. His research interests include Internet of Things, data analysis, and artificial intelligence.

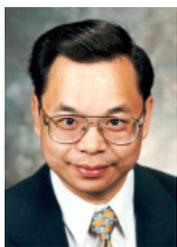

**Simon X. Yang** (Senior Member, IEEE) received the B.Sc. degree in engineering physics from Beijing University, Beijing, China, in 1987, the first of two M.Sc. degrees in biophysics from the Chinese Academy of Sciences, Beijing, China, in 1990, the second M.Sc. degree in electrical engineering from the University of Houston, Houston, TX, in 1996, and the Ph.D. degree in electrical and computer engineering from the University of Alberta, Edmonton, Canada, in 1999.

Dr. Yang is currently a Professor and the Head of the Advanced Robotics and Intelligent Systems Laboratory at the University of Guelph, Guelph, ON, Canada. His research interests include robotics, intelligent systems, sensors and multi-sensor fusion, wireless sensor networks, control systems, machine learning, fuzzy systems, and computational neuroscience.

Prof. Yang has been very active in professional activities. He serves as the Editor-in-Chief of International Journal of Robotics and Automation, and an Associate Editor of IEEE Transactions on Cybernetics, IEEE Transactions on Artificial Intelligence, and several other journals. He has involved in the organization of many international conferences.

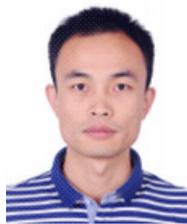

**Tan Guo** (Member, IEEE) received the M.S. degree in signal and information processing and the Ph.D. degree in communication and information systems both from Chongqing University (CQU), Chongqing, China, in 2014 and 2017, respectively. Since 2018, he has been with the School of Communication and Information Engineering, Chongqing University of Posts and Telecommunications (CQUPT), Chongqing, China. He is currently a PostDoctoral Fellow with The Macau University of Science and Technology, Taipa, Macao, China. He is recipient of the Macao Young Scholars Program and the Outstanding Chinese and Foreign Youth Exchange Program of China Association of Science and Technology (CAST). His current research interests include computer vision, pattern recognition, and machine learning.

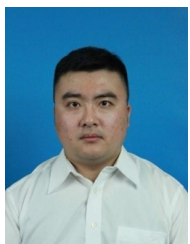

**Qiaozhi Hua** received the M.E. and Ph.D. degrees from the Graduate School of Global Information and Telecommunication Studies, School of Fundamental Science and Engineering, Waseda University, Tokyo, Japan, in 2015 and 2019, respectively. He is currently a University Lecturer at Computer School of Hubei University of Arts and Science, China. His research interests include information-centric networking, Internet of Things, artificial intelligence, C-V2X, edge computing, and information security.

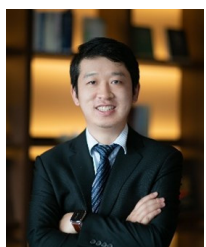

**Keping Yu** (Member, IEEE) received the M.E. and Ph.D. degrees from the Graduate School of Global Information and Telecommunication Studies, Waseda University, Tokyo, Japan, in 2012 and 2016, respectively. He was a Research Associate, Junior Researcher, Researcher with the Global Information and Telecommunication Institute, Waseda University, from 2015 to 2019, 2019 to 2020, and 2020 to 2022, respectively. He is currently an associate professor at Hosei University and a visiting scientist at the RIKEN Center for Advanced Intelligence Project, Japan.

Dr. Yu has hosted and participated in more than ten projects, is involved in many standardization activities organized by ITU-T and ICNRG of IRTF, and has contributed to ITU-T Standards Y.3071 and Supplement 35. He received the IEEE Outstanding Leadership Award from IEEE BigDataSE 2021, the Best Paper Award from IEEE Consumer Electronics Magazine Award 2022 (1st Place Winner), IEEE ICFTIC 2021, ITU Kaleidoscope 2020, the Student Presentation Award from JSST 2014. He has authored 100+ publications including papers in prestigious journal/conferences such as the IEEE WCM, CM, NetMag, IoTJ, TFS, TII, T-ITS, TVT, TMC, JBHI, TR, TCOM, TNSM, TIM, TNSE, TGCN, TCSS, CEM, IOTM, ICC, GLOBECOM etc. He is an Associate Editor of IEEE Open Journal of Vehicular Technology, Journal of Intelligent Manufacturing, Journal of Circuits, Systems and Computers. He has been a Guest Editor for more than 20 journals such as IEEE TCSS. He served as general co-chair and publicity co-chair of the IEEE VTC2020-Spring 1st EBTSRA workshop, general co-chair of IEEE ICCC2020 2nd EBTSRA workshop, general co-chair of IEEE TrustCom2021 3nd EBTSRA workshop, session chair of IEEE ICCC2020, ITU Kaleidoscope 2016. His research interests include smart grids, information-centric networking, the Internet of Things, artificial intelligence, blockchain, and information security.

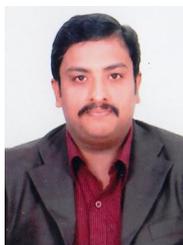

**Arun Kumar Sangaiah** (Member, IEEE) is currently a Professor and engaged as a supervisor of Ph.D and Master Scholars in the National Yunlin University of Science and Technology, Taiwan. He has visited many research centres and universities in China, Japan and South Korea for join collaboration towards research projects and publications. Further, he has contributed more than 300 SCI journal papers, 8 books, 1 patent, 3 projects, among one funded by Ministry of IT of India and few international projects (CAS, Guangdong Research fund, Australian Research Council) cost worth of 50000 USD. Dr. Sangaiah has received many awards, Clarivate Highly Cited Researcher award, Yushan Young Scholar, PIFI-CAS overseas fellowship, Top-10 outstanding researcher, CSI significant Contributor and etc. Also, he is responsible for Editor-in-Chief, book series editor and Associate Editor of various reputed ISI journals.